# The concordance of field-normalized scores based on *Web of Science* and *Microsoft Academic* data: A case study in computer sciences

Thomas Scheidsteger[*], Robin Haunschild[*], Sven Hug[**] and Lutz Bornmann[***]

[*] *T.Scheidsteger@fkf.mpg.de, R.Haunschild@fkf.mpg.de*
Max Planck Institute for Solid State Research, Heisenbergstr. 1, Stuttgart, 70569 (Germany)

[**] *Sven.Hug@evaluation.uzh.ch*
Evaluation Office, University of Zurich, Mühlegasse 21, Zurich, 8001 (Switzerland)

[***] *Bornmann@gv.mpg.de*
Division for Science and Innovation Studies, Administrative Headquarters of the Max Planck Society, Hofgartenstr.8, Munich, 80539 (Germany)

**Introduction**

Microsoft Academic (MA) is a promising new data source for evaluative bibliometrics due to its size and functionality (Harzing & Alakangas, 2017b; Hug & Brandle, 2017; Hug, Ochsner, & Brandle, 2017). Several studies examined the usefulness of MA for citation analysis (Harzing & Alakangas, 2017a; Haunschild, Hug, Brandle, & Bornmann, 2018; Kousha, Thelwall, & Abdoli, 2018; Thelwall, 2017, 2018a; Thelwall, 2018b). The studies generally found that citation counts in MA are similar to those in Scopus, higher than in the Web of Science (WoS), and lower than in Google Scholar. However, these studies – except for Hug, et al. (2017) – did not field-normalize citation counts. Field-normalization is regarded as one of the hallmarks of evaluative bibliometrics in cross-field comparisons. For example, Schubert and Braun (1996) pointed out that "mere publication or citation counts are completely inadequate measures of scientific merit; they can be used for evaluative purposes only after proper standardization or normalization" (p. 311).

Hence, for the purpose of evaluative bibliometrics it is crucial to know, (1) if citation counts from any new bibliometric database, such as MA, could be normalized and (2) whether the normalized scores agree to the scores calculated on the basis of established databases. As the case of MA's closest competitor, Google Scholar, has shown, normalization can be time-consuming, but results in encouraging findings (Bornmann, Thor, Marx, & Schier, 2016; Prins, Costas, van Leeuwen, & Wouters, 2016). In a first attempt, Hug, et al. (2017) calculated normalized scores with data from MA. They normalized the citation impact of papers published by three scholars in one journal across different publication years. Hug, et al. (2017) concluded that normalization based on journals and publication years is easy to perform with MA data. However, they also concluded that the field attributes provided by MA represent a challenge for field-normalization as the attributes are dynamic, fine-grained and field hierarchies are somewhat incoherent.

In the present study, we attempt to tackle the issue of field-normalization for the first time in MA and with a large dataset. We analyze the publications of an anonymous computer science institute with a field-normalized indicator in MA and WoS, thereby assessing the convergent validity of MA. We would like to find an answer on the question whether we receive the same or different normalized scores using WoS or MA data. MA data seems suitable for field-normalization, if we receive similar scores for the institutional papers based on WoS and MA data.

**Methods**
According to the website of the institute, the institute has published 2157 papers between 2005 and 2010.

*Dataset from WoS*
The WoS datasets available in our in-house database developed and maintained by the Max Planck Digital Library (MPDL, Munich) and derived from the Science Citation Index Expanded (SCI-E), Social Sciences Citation Index (SSCI), Arts and Humanities Citation Index (AHCI) provided by Clarivate Analytics (Philadelphia, Pennsylvania, USA) contain disambiguated and unified address information for German research institutes and universities from the Competence Centre for Bibliometrics (CCB, http://www.bibliometrie.info). 1141 papers (52.9%) from the institute were found from the CCB data alone. We performed an address search and found 51 further papers. All of these 1192 papers (55.3%) have at least one WoS subject category, which we used for field-normalization.

*Dataset from MA*
We downloaded the MA datasets from https://aminer.org/open-academic-graph ("MAG papers") on August 15, 2017 (Sinha et al., 2015; Tang et al., 2008). The datasets were imported and processed in our locally maintained database at the Max Planck Institute for Solid State Research (Stuttgart). The datasets were refined to the papers of the research institute by an address search in which 14 different address variants of the institute were considered. Afterwards, 13 false positive papers were found by manual inspection and removed from the dataset of the final papers. In total, we found 2131 papers (98.8%) from the institute. The MA data contain fields of study (FoS) with a four-level hierarchy (L0, L1, L2, and L3). The hierarchical structure used in this study was downloaded from https://academicgraph.blob.core.windows.net/graph/index.html on 02 February, 2016. In principle, fields for normalization can be identified at different hierarchical levels (Waltman & van Eck, in press). The FoS are assigned algorithmically on a paper basis. There are 19 different L0 FoS, 290 different L1 FoS, 1490 different L2 FoS, and 49531 different L3 FoS. We used the L1 FoS as a compromise between granularity of the field classification and significantly populated combinations of FoS and publication year. Furthermore, the 290 different L1 FoS can be expected to provide a similar granularity as the 262 different WoS subject categories. 1714 papers of the institutional paper set were assigned to at least one L1 FoS.

*Affiliation check*
1379 papers (64.7%) from the institute have a DOI in the MA database while only 622 (28.8%) of those papers have a DOI in WoS. 442 papers (20.5%) could be matched via their DOI and this set has been used for our analysis. From a randomly chosen sample of 10% of this set no single paper is incorrectly affiliated. In an analogous sample of the 1699 unmatched papers, we found only 15 papers (0.9%) incorrectly affiliated.

*Normalized citation counts*
Citations were counted in both databases until the end of 2016. Different field-normalization approaches were reviewed by Waltman (2016). The normalized citation score (NCS) (Waltman, van Eck, van Leeuwen, Visser, & van Raan, 2011) is still one of the most popular approaches (van Wijk & Costas-Comesaña, 2012). Therefore, we use the NCS in our study. The citation count of each paper is divided by the average citation count of similar papers. Similar papers are usually defined as papers from the same scientific field and publication year. In MA, we use the L1 FoS and in WoS the WoS subject categories as scientific fields. The NCS is formally defined as

$$NCS = \frac{c_i}{e_i}$$

where $c_i$ is the citation count of a focal paper, and $e_i$ is the corresponding citation rate in the scientific field and publication year (Lundberg, 2007; Rehn, Kronman, & Wadskog, 2007; Waltman, et al., 2011). In the case of multiple scientific field assignments per paper, we calculate the arithmetic average over the multiple NCSs to obtain a single NCS per paper (Haunschild & Bornmann, 2016). The NCS values are named $NCS_{MA}$ and $NCS_{WoS}$, respectively. With $NCS_{MA}$ and $NCS_{WoS}$, we compare in this study an indicator based on a publication-based classification ($NCS_{MA}$) with an indicator based on a journal-based classification ($NCS_{WoS}$).

**Results**

The obvious assumption of a close linear relationship between $NCS_{MA}$ and $NCS_{WoS}$ is confirmed by a Pearson correlation coefficient of $r_p=0.87$ (Spearman correlation coefficient, $r_s=0.84$). The scatterplot in Figure 1 together with a linear regression fit curve gives an impression of the spread of the different institutional $NCS_{MA}$ and $NCS_{WoS}$. Despite a positive relationship between $NCS_{MA}$ and $NCS_{WoS}$, differences in scores are visible especially in the area of high impact scores. Since we are not only interested in the correlation, but also in the reproducibility of both scores (i.e., the concordance), we additionally calculated Lin's concordance correlation coefficient (Lin, 1989, 2000; Liu, 2016) for agreement on a continuous measure. The concordance amounts to $r_{ccc}=0.69$ [0.66, 0.72], which – according to Koch and Sporl (2007) – indicates a "strong" agreement (0.61-0.80). Thus, it seems that both NCS receive similar citation impact results.

Figure 1. Scatterplot of $NCS_{MA}$ and $NCS_{WoS}$. The formula of the linear regression fit is given in the lower right corner

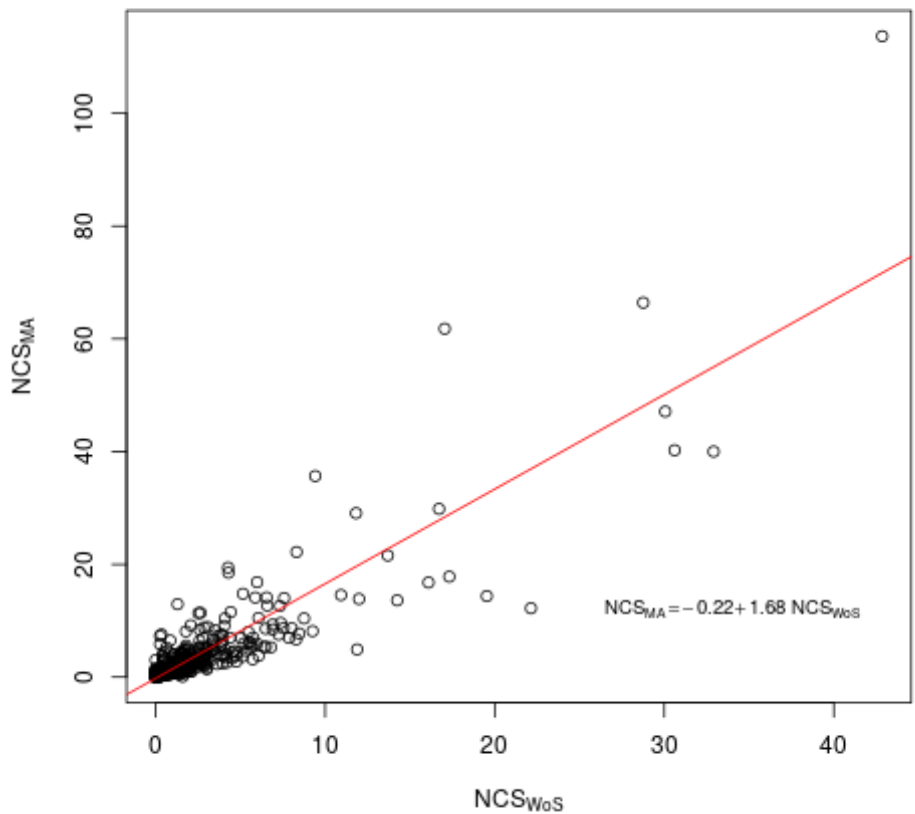

Figure 2 shows the mean $NCS_{MA}$ and $NCS_{WoS}$ with confidence intervals (CIs) (Cumming, 2012). We choose the "paired design" for calculating the CIs, because we have two measures ($NCS_{MA}$ and $NCS_{WoS}$) of the same papers, which results in a reduced CI for the mean difference. The CIs for the two scores do not overlap, so the institute's papers receive a higher $NCS_{MA}$ than $NCS_{WoS}$. Figure 2 also presents the difference between $NCS_{MA}$ and $NCS_{WoS}$ with CIs. As the results show, the difference amounts to 1.3 to 1.7 (with 95% probability).

This is probably due to systematically lower field-specific citation rates $e_i$ for $NCS_{MA}$ than for $NCS_{WoS}$ because MA also includes document types that are usually much less cited than journal articles, e.g. conference papers. As a systematic study of the validity of this claim is currently not feasible because of the unreliability of the document type information given in MA, we manually examined random samples (i.e., 10% of all papers with a DOI and 10% of the DOI-matched subset) to compare true and MA-assigned document types. The first sample contains 52% conference papers (16% in MA), 4% book chapters (0% in MA) and only 44% journal articles (44% in MA) and the second sample comprises only 9% conference papers (5% in MA) and 91% journal articles (89% in MA). This at least gives a strong indication for the prevalence of less cited document types in MA and consequently lower averaged citations. A similar effect can be attributed to the broader coverage of non-English publications in MA. Inside the FoS that are connected with our paper set, about two thirds (62.2%) are written in English.

Figure 2. Mean $NCS_{MA}$ and $NCS_{WoS}$ and their difference – plotted on a floating axis on the right – each with confidence intervals (95%)

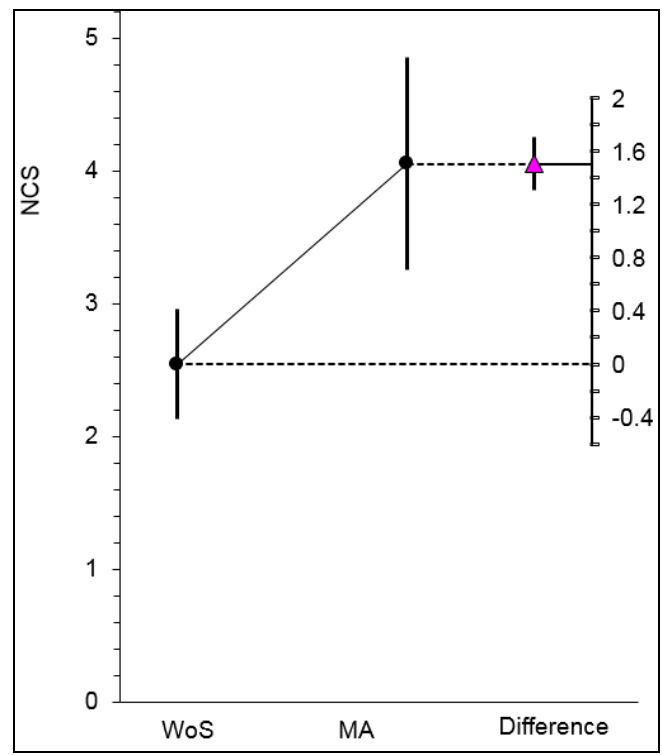

We are additionally interested in the agreement between $NCS_{MA}$ and $NCS_{WoS}$ in different impact classes: do we see the same or different levels of agreement in low or high impact classes, respectively? We used the Characteristic Scores and Scales (CSS) method – proposed by Glänzel, Debackere, and Thijs (2016) – to separate the papers of the institute in different impact classes. For each NCS separately, the classes are obtained by dividing the whole set at the mean value, taking the one class to be all values below the mean, and repeating the division in the set above the respective mean. After three iterations we got four impact classes, which we labeled, following Glänzel, et al. (2016), "poorly cited", "fairly cited", "remarkably cited", and "outstandingly cited". Poorly cited papers are below the average impact of all papers; the other three classes are above this average and further differentiate the papers in the high impact sectors.

The two NCSs are compared in a 4x4 contingency table. The cells in the diagonal enumerate the papers, which have been assigned to the same CSS class for both NCSs. The share of papers assigned in agreement is 81%. Only for one single paper (0.2%) the respective NCS values are more than one class apart. It seems that the concordance between $NCS_{MA}$ and $NCS_{WoS}$ is similarly given in the low as well as in the high impact classes.

Table 1. 4x4 contingency table for the agreement of $NCS_{MA}$ and $NCS_{WoS}$

|  |  | MA | | | |
| --- | --- | --- | --- | --- | --- |
|  |  | poorly cited | fairly cited | remarkably cited | outstandingly cited |
| WoS | poorly cited | 291 | 23 | 1 | 0 |
|  | fairly cited | 32 | 50 | 8 | 0 |
|  | remarkably cited | 0 | 13 | 7 | 2 |
|  | outstandingly cited | 0 | 0 | 4 | 7 |

Based on the values in the 4x4 contingency table, we calculated Cohen's kappa coefficient as a measure of agreement. The coefficient takes into account the possibility of agreement occurring by chance (Gwet, 2014). According to the guidelines of Landis and Koch (1977) for interpreting kappa coefficients, with k = 0.56 the agreement between $NCS_{MA}$ and $NCS_{WoS}$ is "moderate".

**Discussion**
Field-normalization is an important issue in research evaluation, since many bibliometric studies undertake cross-field comparisons (Wilsdon et al., 2015). "Field normalization of scientometric indicators is motivated by the idea that differences between fields lead to distortions in scientometric indicators. One may think of this in terms of signal and noise. Scientometric indicators provide a signal of concepts such as productivity or scientific impact, but they are also affected by noise. This noise may partly be due to differences between fields, for instance differences in publication, collaboration, and citation practices. Field normalization aims to remove this noise while maintaining the signal" (Waltman & van Eck, in press).

The use of MA data for field-normalization has several advantages: (1) a multi-disciplinary classification system is available, (2) the classifications are assigned on the single-paper level (and not on the journal level), and (3) MA covers not only journal papers but also papers with other document types. Thus, field-normalization is in principle possible for all document types (and not only for journal papers as with the WoS).

In this study, we focus on journal papers only, since we compared field-normalized scores based on WoS ($NCS_{WoS}$) with scores based on MA ($NCS_{MA}$). For the comparison, we used the papers of an anonymous institute as an example. The results show a substantial correlation of both scores – based on Pearson and Spearman correlation coefficients ($r_p$ and $r_s$ > 0.8). The concordance given by $r_{ccc}$~0.7 is smaller, but still substantial, displaying a high similarity of both scores. But the non-overlapping confidence intervals of the averages of both scores indicate a statistically significantly higher impact of the paper set in MA (between 1.3 and 1.7) We divided the paper set into four impact classes according to the CSS method and found a similar level of agreement of more than 80% for both scores in all classes. The chance-corrected agreement can be interpreted as "moderate".

Concerning our research questions, our results show that it is possible to calculate field-normalized citations scores from L1 FoS in the MA database, which are in a good agreement with the corresponding scores based on WoS subject categories. Thus, the conclusion of Hug, et al. (2017) – that field-normalization in MA is challenging and perhaps even unfeasible – might have been overly pessimistic. But we need to be aware of some limitations of our approach: we only considered computer science papers, which receive a much broader coverage in MA than in WoS, as well as only those 20% of the institute's papers with a DOI in both databases. These caveats already suggest further studies including the other three FoS levels in MA, other categorization schemes and field coverages as available, for example in Scopus, and the document type.